\documentclass[twocolumn,a4paper,preprintnumbers,aps,prd]{revtex4} 
\usepackage{epsfig}
\usepackage{axodraw}
\usepackage{amsmath}
\usepackage{amssymb}
\usepackage{amsfonts}
\usepackage{graphicx}
\usepackage{graphics,subfigure}
\usepackage{float}
\usepackage{booktabs}
\usepackage{color}
\usepackage{rotating}
\usepackage{atlasphysics}
\usepackage{multirow}
\usepackage{psfrag}
\usepackage{ulem}

\newbox\charbox
\newbox\slabox
\def\s#1{{      
    \setbox\charbox=\hbox{$#1$}
    \setbox\slabox=\hbox{$/$}
    \dimen\charbox=\ht\slabox
    \advance\dimen\charbox by -\dp\slabox
    \advance\dimen\charbox by -\ht\charbox
    \advance\dimen\charbox by \dp\charbox
    \divide\dimen\charbox by 2
    \raise-\dimen\charbox\hbox to \wd\charbox{\hss/\hss}
    \llap{$#1$}
}}

\newcommand{\sw}[1][{}]{\ensuremath{s^{#1}_\mathrm{w}}}
\newcommand{\cw}[1][{}]{\ensuremath{c^{#1}_\mathrm{w}}}

\newcommand{\xo}{\ifmmode {\chi}_1^0\else${\chi}_1^0$\fi}
\newcommand{\sle}{\ifmmode\tilde{\ell}\else$\tilde{l}$\fi}
\newcommand{\selr}{\ifmmode\tilde{e}_R\else$\tilde{e}_R$\fi}

\newcommand{\newc}{\newcommand}
\newc{\ssup}{\tilde{u}}
\newc{\ssdown}{\tilde{d}}
\newc{\ssstrange}{\tilde{s}}
\newc{\sscharm}{\tilde{c}}
\newc{\sstop}{\tilde{t}}
\newc{\ssbottom}{\tilde{b}}
\newc{\sse}{\tilde{e}}
\newc{\ssmu}{\tilde{\mu}}
\newc{\sstau}{\tilde{\tau}}
\newc{\ssnue}{\tilde{\nu}_{e}}
\newc{\ssnumu}{{\tilde{\nu}_{\mu}}}
\newc{\ssnutau}{{\tilde{\nu}_{\tau}}}
\newc{\ssbnue}{\tilde{\nu}_{e}^*}
\newc{\ssbnumu}{\tilde{\nu}_{\mu}^*}
\newc{\ssbnutau}{\tilde{\nu}_{\tau}^*}

\def\rpv{\mbox{\ensuremath{\, \slash\kern-.6emR_{p}}}}

\begin{document}
 
\title{Measuring a Light Neutralino Mass at the ILC: Testing the MSSM
Neutralino Cold Dark Matter Model}

\author{J.~A.~Conley} \email[]{conley@th.physik.uni-bonn.de}
\affiliation{Physikalisches Institut and Bethe Center for Theoretical
Physics, Universit\"at Bonn, Nu{\ss}allee 12, 53115 Bonn, Germany}

\author{H.~K.~Dreiner}
\email[]{dreiner@th.physik.uni-bonn.de}
\affiliation{Physikalisches Institut and Bethe Center for Theoretical Physics, 
Universit\"at Bonn, Nu{\ss}allee 12, 53115 Bonn, Germany}
\affiliation{SCIPP, University of California, Santa Cruz, CA 95064}

\author{P.~Wienemann} 
\email[]{wienemann@physik.uni-bonn.de}
\affiliation{Physikalisches Institut, Universit\"at Bonn, Nu{\ss}allee
12, 53115 Bonn, Germany}

\begin{abstract}
The LEP experiments give a lower bound on the neutralino mass of about
46 GeV which, however, relies on a supersymmetric grand unification
relation.  Dropping this assumption, the experimental lower bound on
the neutralino mass vanishes completely. Recent analyses suggest,
however, that in the minimal supersymmetric standard model (MSSM), a
light neutralino dark matter candidate has a lower bound on its mass
of about 7 GeV. In light of this, we investigate the mass sensitivity
at the ILC for very light neutralinos. We study slepton pair
production, followed by the decay of the sleptons to a lepton and the
lightest neutralino. We find that the mass measurement accuracy for a
few-GeV neutralino is around 2 GeV, or even less if the relevant
slepton is sufficiently light.  We thus conclude that the ILC can help
verify or falsify the MSSM neutralino cold dark matter model even for
very light neutralinos.
\end{abstract}

\preprint{BONN-TH-2010-12}
\preprint{BONN-HE-2010-01}

\maketitle

\section{Introduction}
The supersymmetric Standard Model (SSM)
\cite{Nilles:1983ge,Martin:1997ns} is a well motivated extension of
the Standard Model of particle physics which solves the hierarchy
problem between the weak scale and the Planck scale
\cite{Gildener:1976ih,Veltman:1980mj}. In order to guarantee a stable
proton, usually a discrete symmetry beyond the SM gauge symmetries
is imposed which prohibits the baryon-- and lepton--number violating
terms in the superpotential: R--parity \cite{Farrar:1978xj}, proton
hexality \cite{Dreiner:2005rd}, or a $\mathbf{Z}_4$ R--symmetry
\cite{Lee:2010gv}. This is then called the minimal supersymmetric 
Standard Model (MSSM). The discrete symmetry furthermore guarantees
that the lightest supersymmetric particle (LSP) is stable and thus a
dark matter candidate
\cite{Pagels:1981ke,Ellis:1983ew,Jungman:1995df,Drees:1996pk}. If it
is to constitute the entire dark matter in the universe it must be
electrically and color neutral \cite{Ellis:1983ew}. Here we focus on
the lightest neutralino $\chi^0_1$ as the LSP. In order to avoid
overclosure of the universe, the neutralino must be either very light
$M_{\chi^0_1}<\mathcal{O}(1\,\mathrm{eV})$ (Cowsik--McLelland bound)
\cite{Cowsik:1972gh,Dreiner:2009ic} or heavy $M_{\chi^0_1}>
\mathcal{O}(10\,\mathrm{GeV})$ (Lee--Weinberg bound) \cite{Lee:1977ua}.
We shall make this latter lower bound more precise
\cite{Hooper:2002nq,Belanger:2002nr,Bottino:2002ry,Bottino:2003cz,Bottino:2003iu,Bottino:2004qi,Bottino:2007qg}.
In this paper we are interested in how a neutralino mass close to the
Lee--Weinberg bound could be measured at the ILC. This is potentially a
stringent test of a MSSM light dark matter model.

Within the MSSM the spin--1/2 superpartners of the hypercharge $B$
boson, the neutral SU(2) $W$ boson and the two CP--even neutral Higgs
bosons mix after electroweak symmetry breaking. The resulting four
mass eigenstates are the neutralinos and are denoted $\chi^0_i$, $i=1
,\ldots,4$. The masses are ordered $M_{\chi^0_1}\!<\!\ldots\!<\!M_{
\chi^0_4}$. If produced at colliders, the lightest neutralino behaves
like a heavy stable neutrino and escapes detection. The spin--1/2
superpartners of the charged SU(2) $W$ boson and of the charged Higgs
boson also mix after electroweak symmetry breaking. The resulting mass
eigenstates are the charginos and denoted $\chi^\pm_{i=1,2}$, with
ordered masses.  See Appendix A for details.

The current Particle Data Group (PDG) mass bound from LEP on the
lightest neutralino is
\cite{Abdallah:2003xe,Heister:2003zk,Abbiendi:2003sc,Amsler:2008zzb}
\begin{equation}
M_{\chi^0_1}\,>\, 46\gev\,.
\label{lep-bound}
\end{equation}
This bound is obtained by searching for charginos and thus setting a
bound on the SU(2) gaugino mass $M_2$ and the Higgs mixing parameter
$\mu$. Using the supersymmetric grand unified theory (SUSY GUT)
relation between $M_2$ and the U(1)$_Y$ gaugino mass term $M_1$
\begin{equation}
M_1=\frac{5}{3}\tan^2\theta_\mathrm{w}\, M_2\,,
\label{susy-gut}
\end{equation}
the chargino search can be translated into a bound on $M_1$. The
neutralino mass matrix is computed for all allowed values of the
supersymmetric parameters, taking into account Eq.~(\ref{susy-gut}),
as well as the lower bound on the ratio of the Higgs vacuum
expectation values, $\tan\beta\gtrsim2$, 
from the LEP Higgs search \cite{Amsler:2008zzb}. Then one obtains as
the lowest possible neutralino mass the bound in
Eq.~(\ref{lep-bound}).  If, however, the assumption
Eq.~(\ref{susy-gut}) is dropped, there is no lower laboratory or
astrophysical bound on the neutralino mass
\cite{Choudhury:1999tn,Kachelriess:2000dz,Dedes:2001zia,Gogoladze:2002xp,Dreiner:2003wh,Dreiner:2007vm,Dreiner:2009ic,Dreiner:2009er}. 
Even a massless neutralino is allowed. This is now included in the PDG
as a comment \cite{Nakamura:2010xx}. There is however a cosmological
bound which we now discuss.

Relaxing the SUSY GUT assumption in Eq.~(\ref{susy-gut}), it is
possible to derive the Lee--Weinberg lower limit on the mass of the
neutralino LSP, $M^{\rm min}_{\xo}$, in the MSSM with real
parameters. It was first determined for large pseudoscalar Higgs
masses \cite{Hooper:2002nq,Belanger:2002nr}, obtaining $M_{\chi^0_1}^
{\mathrm {min}}=\mathcal{O}(15\,\mathrm{GeV})$. It was subsequently
realized however \cite{Bottino:2002ry,Bottino:2003iu}, that a region
of parameter space exists with a low pseudoscalar Higgs mass and high
$\tan \beta$, in which the neutralino lower mass limit reaches $M^{\rm
min}_{\xo}\approx6\GeV$. This is due to an enhancement in the
neutralino annihilation cross section from annihilation to b-quarks
via Higgs bosons, which keeps the predicted relic density below the
observed limits. This was confirmed in Ref.~\cite{Belanger:2003wb}.
There it was furthermore noted that this area of parameter space would
be testable at the Tevatron, for example, with the Higgs search
results in which the Higgs is produced in association with a b-quark,
as well as via the $B_s\to\mu^+\mu^-$ limit. See also the more recent
work in
Ref.~\cite{Bottino:2008xc,Niro:2009mw,Badin:2009cf,Feldman:2010ke}.
In a very recent paper \cite{Fornengo:2010mk}, the authors argue that
these constraints have a relatively minor impact on the light neutralino
parameter space of the MSSM \footnote{Note, however that in an earlier work
\cite{Vasquez:2010ru},  this region of parameter space is also examined, and
there it is argued that 
the Tevatron Higgs searches, the most recent $B_s\to\mu^+\mu^-$ limit, and
dark matter direct detection null results together rule out
neutralinos lighter than 28 GeV in the MSSM.}, and that the lower bound is
\begin{equation}
  M^{\rm min}_{\xo}\approx 7-8\GeV\,.
  \label{mchi-bound}
\end{equation}

Recently there has also been an increased interest in light dark
matter candidates with a mass of order 5 GeV due to the DAMA/LIBRA
\cite{Bernabei:2010mq} and CoGeNT \cite{Aalseth:2010vx} direct search
results. Ref. \cite{Kuflik:2010ah} suggests that the required scattering 
cross sections in the detectors cannot be obtained within the MSSM,
though in Ref. \cite{Fornengo:2010mk} it is claimed that the current
constraints do in fact allow sufficiently high cross sections for these
experimental hints to be explained by an MSSM neutralino. 
In the NMSSM, it is similarly claimed in Ref.~\cite{Das:2010ww,Gunion:2010dy} that
high cross sections cannot be obtained,
though Refs.~\cite{Draper:2010ew,Vasquez:2010ru} identify regions of parameter
space in this model in which the presence of a light singlet Higgs can lead to large
enough cross sections. The authors of Refs.~\cite{Belikov:2010yi} argue
that solutions also exist for an extended NMSSM. The experimental
results in Refs.~\cite{Bernabei:2010mq,Aalseth:2010vx} have lead to a
flourish of alternative schemes,
e.g.\ \cite{Kim:2009ke,Fitzpatrick:2010em,Andreas:2010dz,Frandsen:2010yj,Alves:2010dd,Feldman:2010wy,Essig:2010ye,Chang:2010yk,Graham:2010ca,Kappl:2010qx,Bae:2010hr}. 
We note, however, that these models are severely constrained by the
CDMS-II \cite{Ahmed:2008eu,Ahmed:2009zw} and XENON10
\cite{Angle:2007uj} and XENON100 data \cite{Aprile:2010um,Baudis:2010tk}.

Given these considerations it is thus of great interest how well the
neutralino mass can be determined in the low-mass region. It is the
goal of this paper to test the sensitivity of the mass measurement at
the ILC. There is the possibility that a neutralino mass will be
measured which is too small to be reconciled with the observed relic
abundance, if the real MSSM and standard cosmology are assumed. Such a
measurement would be striking evidence for non-minimal supersymmetric
models and/or non-standard cosmologies.

The paper is organized as follows.  In section \ref{neutMassMeas}, we
provide an overview of some of the methods that have been suggested to
measure the lightest neutralino mass at the ILC.  We then focus, in
Section \ref{slepPair}, on the neutralino mass measurement that can be
done using slepton pair production, and make a first estimate of the
accuracy of this method for a very light neutralino.  In Section
\ref{simulation}, we describe a simulation of slepton pair
production at the ILC, and use it to make a better determination of
the accuracy attainable for the neutralino mass measurement.  We
summarize and conclude in Section \ref{conclusions}.

\section{Neutralino Mass Measurements}
\label{neutMassMeas}

Several methods have been suggested in the literature to measure the
mass of the lightest neutralino at the ILC
\cite{Martyn:2004ew,Martyn:2004jc,Freitas:2004re,MoortgatPick:2008yt,:2007sg}.
Throughout, the authors have focused on a neutralino heavier than the
LEP bound in Eq.~(\ref{lep-bound}). For example, the widely studied
SPS1a point (without a slope) \cite{Allanach:2002nj} has
$M_{\chi^0_1}=97.1\gev$. The most straightforward method involves
considering slepton pair production, followed by the decay of each
slepton to the lightest neutralino and a charged lepton
\begin{eqnarray} e^+ e^-&\rightarrow& \tilde \ell^- \tilde
\ell^+ \rightarrow \ell^-\ell^++2 \chi^0_1\,,\quad \ell=e,\,\mu\,.
\label{slepton-pair}
\end{eqnarray}
Here the (s)leptons are restricted to the first two generations. The
measurement via the third generation (s)tau is diluted by the
additional decay to the neutrino(s). Measuring the energies of the final
state leptons, one can extract information on the neutralino and
slepton masses. The typical relative precision achieved is in the per
mille range
\cite{Martyn:2004ew,Martyn:2004jc,Freitas:2004re,MoortgatPick:2008yt}.
We go beyond this work and discuss this method in detail
for a very light neutralino.

A second method in the literature is based on the pair production of
the second lightest neutralinos. This is followed by the decay of each
neutralino via a (virtual) slepton to a charged lepton pair and the
lightest neutralino,
\begin{eqnarray}
e^+ e^-&\rightarrow& \chi^0_2 \chi^0_2\rightarrow
(\chi^0_1 \ell_1^+\ell_1^-)\;(\chi^0_1 \ell_2^+\ell_2^-)
\,.
\label{neut-pair}
\end{eqnarray}
where each $\chi_2^0$ decays independently and thus $\ell_1$ need not
equal $\ell_2$. In fact, the case $\ell_1\not=\ell_2$ reduces the
combinatorial uncertainty. In
Refs.~\cite{AguilarSaavedra:2001rg,Djouadi:2007ik} the authors then
propose to measure the di-lepton invariant mass and the di-lepton
energy and to use these to measure the two lightest neutralino masses.

Of necessity these methods also always involve other supersymmetric
particles and their masses. For example, the first method relies on
the production of sleptons. The second method relies on the production
of the second lightest neutralino and then its decay to an
intermediate slepton. Thus both of these methods can be improved by
measuring the corresponding supersymmetric masses directly. For
example the slepton mass can be well determined by an energy scan over
the slepton mass threshold \cite{AguilarSaavedra:2001rg}. Similarly a scan over
the production threshold energy of the process given in
Eq.~(\ref{neut-pair}) gives a tight constraint on the mass
$M_{\chi^0_2}$ \cite{AguilarSaavedra:2001rg}.

\section{Slepton Pair Production and the Neutralino Mass}
\label{slepPair}

In this section, we study the measurement of the lightest neutralino
mass using slepton pair production at the ILC, as shown in
Eq.~(\ref{slepton-pair}).  The slepton decay to a lepton and the
lightest neutralino is a two-body decay. Therefore in the slepton
rest-frame the lepton energy is completely fixed by the slepton and
neutralino mass. Ignoring initial and final state radiation (ISR and
FSR), beamstrahlung, and detector effects for the moment, the slepton
energy is then just the beam energy. Thus the lepton's lab-frame
energy $E_\ell$ is fully determined by the angle $\theta_0$, with
which the slepton emits the lepton in the slepton rest-frame. The
angle is measured with respect to the slepton lab momentum
direction. We then have for the lepton energy
\begin{equation}
E_\ell=\frac{\sqrt{s}}{4}\left(1-\frac{M_{\xo}^2}{M_{\sle}^2}\right)
(1+\beta\cos{\theta_0})\;.\label{slep-energy}
\end{equation}
Here $\beta=\sqrt{1-4m_{\sle}^2/s}$ is the slepton velocity in the lab
frame, $\sqrt{s}/2$ is the beam energy, and $M_{\sle}$ denotes the
slepton mass.  The event distribution of $E_\ell$ is flat between its
maximum $E_{+}$, when $\cos{\theta_0}=1$, and its minimum $E_{-}$,
when $\cos{\theta_0}=-1$.  The equations for $E_{+}$ and $E_{-}$ can
be inverted to find the slepton and neutralino masses squared in terms
of these endpoints,
\begin{equation}
  \label{msl2}
  M_{\sle}^2=s\,\frac{{E_{+}E_{-}}}{(E_{+}+E_{-})^2}\;,
\end{equation}
and
\begin{equation}
  \label{mxo2}
  M_{\xo}^2=M_{\sle}^2\left(1-\frac{E_{+}+E_{-}}{\sqrt{s}/2}\right)\;.
\end{equation}
We have listed the squared formul{\ae} for later use. Taking the
positive square root we then obtain for the masses
\begin{equation}
  \label{msl}
  M_{\sle}=\sqrt{s}\,\frac{\sqrt{E_{+}E_{-}}}{E_{+}+E_{-}}\;,
\end{equation}
and
\begin{equation}
  \label{mxo}
  M_{\xo}=M_{\sle}\sqrt{1-\frac{E_{+}+E_{-}}{\sqrt{s}/2}}\;.
\end{equation}
The sensitivity of the neutralino mass measurement thus depends on the
accuracy with which $E_{\pm}$ can be measured.

Looking at Eq.~(\ref{slep-energy}), it is clear that $E_\pm$ only have
a weak dependence on $M_{\xo}$ for $M_{\xo} \muchless M_{\sle}$. Thus we
would expect the accuracy of the neutralino mass measurement to
deteriorate for sufficiently small neutralino masses.  We set out to
quantify this below.

Limited statistics and detector and beam effects introduce uncertainty
into the endpoint determination.  Nonetheless, for typical slepton and
heavy neutralino masses, the endpoints and the masses can be
determined to sub-GeV accuracy~\cite{Martyn:2004ew}.  For very light
neutralinos, however, even small errors in the endpoint measurements
can lead to a large fractional error in the neutralino mass
determination.

Before studying this issue with a simulation, we can estimate
the mass determination accuracy for a light neutralino by combining
the quoted accuracy from an experimental study by Martyn
\cite{Martyn:2004ew} with a simple error analysis.  Assuming that
$E_+$ and $E_-$ are independent random variables, then from
Eq.~(\ref{mxo}) we can derive
\begin{equation}
\label{dmxo}
\frac{\delta M_{\xo}}{\delta E_{\pm}}=\frac{\delta M_{\sle}}{\delta 
E_{\pm}} \frac{M_{\xo}}{M_{\sle}} - \frac{M_{\sle}^2}{M_{\xo}\sqrt{s}}\,.
\end{equation}
For light neutralinos, the first term in Eq.~(\ref{dmxo}) is
negligible.  The second term dominates and is identical for $E_+$ and
$E_-$, so we can write
\begin{equation}
  \label{dmxoapp}
  \delta M_{\xo}\simeq \frac{M_{\sle}^2}{M_{\xo}\sqrt{s}}
  \sqrt{\delta E_+^2+\delta E_-^2}\,.
\end{equation}
In the simulation we describe below, we consider SUSY
scenarios with $M_{\tilde{e}_R}=100$ and 200~GeV and varying
$M_{\xo}$, a center-of-mass energy $\sqrt{s}=500$~GeV and an
integrated luminosity $\mathcal{L}=250$~fb$^{-1}$.  For illustration,
we here assume these experimental parameters as well as a 2
GeV neutralino mass and 100 GeV selectron mass.
Thus the factor in front of the square root in
Eq.~(\ref{dmxoapp}) is 10.

In Ref.~\cite{Martyn:2004ew}, the error on the endpoint determinations
is given as $\delta E_+=0.11$ GeV and $\delta E_-=0.02$~GeV for a
scenario with $M_{\xo}=93$~GeV, $M_{\sle}=143$~GeV, $\mathcal{L}=200
$~fb$^{-1}$, and $\sqrt{s}=400$~GeV.  In this scenario, because the
neutralino is heavy the two terms in Eq.~(\ref{dmxo}) are comparable,
so we cannot use Eq.~(\ref{dmxoapp}).  Using 
Eq.~(\ref{dmxo}) instead, we obtain $\delta M_{\xo}\simeq 100\,\mathrm{MeV}$,
which agrees exactly with the quoted result of the detailed study in
Ref.~\cite{Martyn:2004ew}.

To translate these into an estimate for $\delta E_{\pm}$ in our
scenario, we need to take into account several modifications. (i)~The
endpoint locations have changed significantly because the slepton and
neutralino masses are different. Therefore also the experimental
energy resolution at the endpoint locations is different. (ii)~The
number of events for slepton pair production is different in our
scenario due to the different masses, center-of-mass energy, and
luminosity, so that the statistical error on the endpoint
determination is different.  Determining the effect of (i) requires
choosing a parametrization of the detector's energy resolution, which
we discuss in the next section and provide in Eqs.~(\ref{pres}) and
(\ref{eres}).  Taking these two factors into account, we can estimate
the ratio of our endpoint energy uncertainty to Martyn's in
Ref.~\cite{Martyn:2004ew}
\begin{equation}
  \frac{\delta E_{\pm}^{\mathrm{us}}}{\delta
    E_{\pm}^{\mathrm{Martyn}}}\simeq
  \frac{\delta E_{\mathrm{exp}}(E=E_{\pm}^{\mathrm{us}})}{\delta
    E_{\mathrm{exp}}(E=E_{\pm}^{\mathrm{Martyn}})}
  \times
  \sqrt{\frac{N_{\mathrm{events}}^{\mathrm{Martyn}}}
    {N_{\mathrm{events}}^{\mathrm{us}}}}\,,
\end{equation}
where we have estimated that the uncertainty in the endpoint 
determination drops with the square root of the number of observed events.
Plugging the relevant numbers into the above expression for $E_+$,
which dominates the error, in fact yields $\frac{\delta
E_+^{\mathrm{us}}}{\delta E_+^{\mathrm{Martyn}}}\simeq 1.3$, since the
increased number of events in our scenario is partially canceled
by the reduced detector resolution at the higher value of $E_+$.

Referring again to Eq.~(\ref{dmxoapp}), we can then estimate that
$\delta M_{\xo}\simeq 1.4$ GeV for our scenario.  In other words, for
a neutralino mass of 2 GeV, the mass can be determined to about 70\%
accuracy.  This suggests that a useful mass measurement can be
performed for very light neutralinos, and in particular in the range
$M_{\xo}\sim 5$ GeV that is particularly interesting for dark matter
phenomenology, sub-GeV accuracy should be possible.

On the other hand, if we carry out the same estimate for a 2~GeV
neutralino and instead a 100~GeV $\tilde{e}_R$, we find that the
factor in front of the square root in Eq.~(\ref{dmxoapp}) is now 40,
and the cross section for selectron pair production is also lower so
that the statistical uncertainty is larger.  In this case we find
$\delta M_{\xo}\simeq 15$ GeV, suggesting that in this case at best an
upper limit on the neutralino mass can be set.

While this simple estimate gives a qualitative illustration of the
difficulty of measuring a light neutralino mass, we would
would like to check it with a more thorough analysis and more precisely quantify the
accuracy possible for a light neutralino mass measurement at the ILC.
We do this in the next section.

\section{Simulation of Neutralino Mass Measurement from Slepton Pair Production}
\label{simulation}

Thanks to the simple kinematics of slepton pair production, it is
possible to estimate the precision for a ${\chi}_1^0$ mass
measurement at the ILC from a rather simple Monte Carlo simulation.
We describe this in the following.

First, the number of produced slepton pairs for a given centre-of-mass
energy $\sqrt{s}$ and luminosity $\mathcal{L}$ is calculated for a
beam polarisation of $(\mathcal{P}_{e^-}, \mathcal{P}_{e^+}) = (+80
\%,-60 \%)$ using the program {\tt SPheno}~\cite{Porod:2003um}, which
implements the cross section formulae from
Refs.~\cite{Bartl:1987zg,Bartl:1997yi,Blochinger:2002zw,Kraml:1999qd}.
This choice of signs for the beam polarization maximizes the
production cross-section. For each event, two lepton energies are
thrown according to a flat probability density distribution between
$E_-$ and $E_+$. In order to take effects caused by beamstrahlung into
account, $E_-$ and $E_+$ are evaluated for each event using the
reduced centre-of-mass energy $\sqrt{s^{\prime}}$ which is thrown
according to the luminosity spectrum computed by GUINEA
PIG~\cite{Schulte:1999tx}. The energy difference $\sqrt{s} -
\sqrt{s^{\prime}}$ is lost in the form of beamstrahlung photons. As a
result the sharp edges in the lepton energy spectrum are smoothed out
a bit.

The resulting lepton energies are subsequently smeared according to
the expected momentum and energy resolution. This
smoothes out the edges even further. For electrons, the minimum of
track momentum resolution and
the energy resolution of the electromagnetic
calorimeter (ECAL) for the considered electron energy is
employed. In the case of muons, the momentum
resolution of the tracking system is always used. For these
quantities, the following parametrizations are used:
\begin{eqnarray}
    \Delta \frac{1}{p_T} & = & 1 \cdot 10^{-4}~\mathrm{GeV}^{-1} 
\qquad \mbox{(tracker),} \label{pres}\\
    \frac{\Delta E}{E} & = & \frac{0.166}{\sqrt{E/\mathrm{GeV}}} 
\oplus 0.011 \quad \mbox{(ECAL).}
\label{eres}
\end{eqnarray}
Any polar angle dependence of the tracker resolution is neglected.
Instead a rather conservative average resolution is applied (compare
e.g.\ Ref.~\cite{ILDLOI}). We checked that the results do not depend
strongly on the assumed tracker resolution since for the considered
SUSY masses, the \xo~mass measurement is dominated by the calorimeter
resolution. The above parametrization of the ECAL resolution which we
employ is the one obtained with a detector prototype in test beam
measurements~\cite{calice:2008ma}. The effects of a limited detector
acceptance, signal selection cuts and inefficiencies in the electron
and muon reconstruction are approximately accounted for by applying an
overall efficiency of 50\%. This roughly corresponds to the values
obtained in Ref.~\cite{Martyn:2004ew} using a more detailed
simulation. This more detailed study also showed that background rates
are rather small~\cite{AguilarSaavedra:2001rg}. Therefore, outside of
this overall efficiency, we neglect backgrounds completely in our
study.

The edge positions of the lepton spectrum obtained in the described
way are finally fitted using an unbinned likelihood fit. The fitted
shapes are
\begin{equation}
    f_-(E) = \left\{ \begin{array}{r@{\quad:\quad}l} \frac{1}{2}
      \left[ \mathrm{erf}\left(
        \frac{E-\hat{E}_-}{\sqrt{2}\sigma_1^-}\right) + 1\right] & E < \hat{E}_-
      \\ \frac{1}{2} \left[ \mathrm{erf}\left(
        \frac{E-\hat{E}_-}{\sqrt{2}\sigma_2^-}\right) + 1\right] & E \ge
      \hat{E}_- \end{array} \right.
\label{eq:fm}
\end{equation}
for $E_-$ and
\begin{equation}
    f_+(E) = \left\{ \begin{array}{r@{\quad:\quad}l} \frac{1}{2}
      \mathrm{erfc}\left(
      \frac{E-\hat{E}_+}{\sqrt{2}\sigma_1^+}\right) & E < \hat{E}_+
      \\ \frac{1}{2} \mathrm{erfc}\left(
      \frac{E-\hat{E}_+}{\sqrt{2}\sigma_2^+}\right) & E \ge
      \hat{E}_+ \end{array} \right.
\label{eq:fp}
\end{equation}
for $E_+$. If one chooses $\sigma_1^{\pm} = \sigma_2^{\pm}$,
Eqs~(\ref{eq:fm}) and~(\ref{eq:fp}) are the results of a convolution
of an upward and a downward step function with a Gaussian. Between the
nominal edge positions $E_-$ and $E_+$ the shape of the lepton energy
spectrum is influenced by beamstrahlung and energy/momentum
resolution, whereas outside the nominal edge positions, the shape is
only determined by the energy/momentum resolution. For this reason
$\sigma_1^{\pm}$ and $\sigma_2^{\pm}$ are treated as separate
parameters in the fit. The fitted values of the parameters $\hat{E}_-$
and $\hat{E}_+$ do not in general coincide with the values of $E_-$
and $E_+$. The reason is that the asymmetric shape of the
beamstrahlung energy spectrum leads to a certain offset. To correct
for this bias, a Monte Carlo based calibration procedure is used.

The uncertainty on the edge positions, and thus the masses, is
determined by creating toy Monte Carlo datasets.  For each toy data
set, the fitted and corrected values of $E_+$ and $E_-$ can be
converted into the squared neutralino mass using Eq.~(\ref{mxo2}).
The distribution of $m_{\xo}^2$ from the ensemble toy data sets is
approximately Gaussian, and we use the width of this distribution to
determine the uncertainty on the mass measurement.  For low neutralino
masses, the distribution can have support in the unphysical region
where $m_{\xo}^2<0$. To account for this, we used the Feldman-Cousins
method~\cite{Feldman:1997qc} to have a smooth transition between a
mass measurement, which is possible for heavier neutralinos, and an
upper bound, which is necessary for very light neutralinos.

Using the described procedure, we agree within roughly 30\% with the
results in Refs.~\cite{AguilarSaavedra:2001rg,Martyn:2004ew,Martyn:2004jc},
which were obtained for $M_{\xo}=71.9$, 96, and 135 GeV, respectively.
Therefore we assign a systematic uncertainty of 30\% to our
results due to the simplifications of our simulation.

\begin{figure}
\includegraphics[width=1.0\columnwidth]{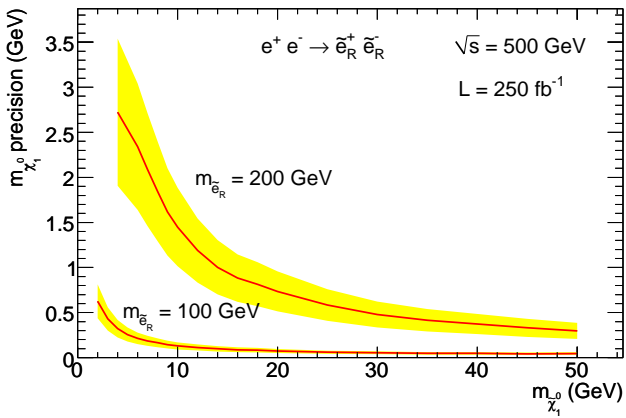}
\caption{Estimated precision of the $\chi_1^0$ mass measurement from
  $\tilde{e}_R\tilde{e}_R$ production as function of the
  ${\chi}_1^0$ mass for $\tilde{e}_R$ masses of 100~GeV and
  200~GeV. The yellow bands represent the estimated uncertainty of
  30\% related to the simplifications of the Monte Carlo
  simulation used. The assumed centre-of-mass energy is $\sqrt{s} =
  500$~GeV, the integrated luminosity $\mathcal{L} = 250$~fb$^{-1}$
  and the beam polarization $(\mathcal{P}_{e^-}, \mathcal{P}_{e^+}) =
  (+80 \%,-60 \%)$.}
\label{fig:chi10massresolution}
\end{figure}

Our estimate of the precision of the ${\chi}_1^0$ mass measurement
from $\tilde{e}_R\tilde{e}_R$ production at the ILC is shown in
Fig.~\ref{fig:chi10massresolution} as a function of the $\chi_1^0$
mass. The assumed luminosity is 250~fb$^{-1}$ at a centre-of-mass
energy of $\sqrt{s} = 500$~GeV with a beam polarization of
$(\mathcal{P}_{e^-}, \mathcal{P}_{e^+}) = (+80 \%,-60
\%)$. Even for $\chi_1^0$ masses as small as 2~GeV, a
precision on the $\chi_1^0$ mass measurement of $\approx
0.6\,\gev$ can be achieved for $M_{\selr}=100\,\gev$.

We find that below about 2 (4)~GeV for a 100 (200)~GeV selectron, a mass
measurement is no longer possible and we can only set an upper bound on the
neutralino mass.  For example, for $m_{\chi_1^0} = 1$~GeV, the 95 \% CL upper
limits are ~2.5~GeV (7.6~GeV) for a selectron mass of 100~GeV (200~GeV).

We note that the precision of the mass measurement that we obtain in
this simulation is roughly a factor of two better than the rough
estimate of the precision made in Section \ref{slepPair}.  This is due
to the simplistic scaling assumptions made there, namely that the
endpoint energy determination accuracies scale like $1/\sqrt{N}$ as
the number of events changes, and linearly with the detector
resolution as the endpoint energy changes.  Using dedicated
simulations we find some deviation from this simple scaling which is
due in a large part to the effect of beamstrahlung.  This more
realistic scaling can account for the discrepancy between our estimate
and simulation results.

Combining the results from $\tilde{\mu}_R\tilde{\mu}_R$ production
with those from $\tilde{e}_R\tilde{e}_R$ production does not lead to a
sizable improvement of the obtained precision. The reason is the
significantly higher cross-section for $\tilde{e}_R\tilde{e}_R$
production due to the additional $t$-channel contribution, which is
especially important for low neutralino masses, leading to a factor of
2 to 3 weaker constraints from $\tilde{\mu}_R\tilde{\mu}_R$
production.

\section{Summary and Conclusion}
\label{conclusions}

A light neutralino in the several-GeV mass range is currently of
special phenomenological interest.  Recent dark matter direct
detection experiments hint at the possible existence of such a light
particle.  On the other hand, recent phenomenological analyses claim
that an MSSM light neutralino dark matter candidate has a lower bound
on its mass around 7~GeV.

If a light neutralino exists, it would therefore be extremely important to
obtain an accurate determination of its mass.  Techniques for measuring
neutralino masses at the ILC have been developed and shown to have
extraordinary precision for the more conventional 50--100 GeV range.  These
techniques, however, have not been studied for much lighter neutralinos.

In this paper, we have studied one of these techniques---measuring the
lepton energy spectrum in slepton pair production events---and
determined its usefulness for the measurement of very light neutralino
masses.  We showed with a simulation that this technique continues to
have useful accuracy for a neutralino with a mass as low as a few GeV.
For example, we showed that it is possible to measure the mass of even
a 2~GeV neutralino to sub-GeV accuracy if the mass of the right-handed
selectron is 100~GeV.  For a 200~GeV selectron, the precision is
about~2.7 GeV for a 4~GeV neutralino.  For even lighter neutralinos,
we showed that this method can give an 95 \% CL upper bound of 2.5
(7.6)~GeV for a 100 (200)~GeV selectron.

Such mass measurements at the ILC will thus be indispensable in
testing the MSSM thermal cold dark matter picture if
a very light neutralino exists.

\appendix

\section{Chargino and Neutralino Mixing}
Here we summarize the mixing of the electroweak gauginos and
Higgsinos, which we use in the paper. The spin-1/2 superpartners of
the $W^\pm$ gauge bosons and the scalar charged Higgs field, $H^\pm$
mix after electroweak symmetry breaking.  The resulting mixing matrix
in the wino, Higgsino basis is given by \cite{Bartl:1985fk}
\begin{equation}
\begin{pmatrix}
M_2 & \sqrt{2}M_W\sin\beta\\  
\sqrt{2}M_W\cos\beta   & \mu
\end{pmatrix}\,.
\end{equation}
Here $M_2$ is the SU(2) soft breaking gaugino mass. $\mu$ is the
supersymmetric Higgs mixing parameter, $\tan\beta$ is the ratio of the
vacuum expectation values of the two Higgs doublets and
$M_W$ is the mass of the $W$ boson.

Similarly the spin-1/2 superpartners of the $W^0$ and $B$ gauge
bosons as well as of the two CP-even neutral Higgs mix after
electroweak symmetry breaking. The $4\times4$ mixing matrix is given
in the bino, wino, Higgsino basis by \cite{Bartl:1986hp}
\begin{equation}
\begin{pmatrix}
M_1 & 0   & - \MZ \sw c_\beta & \phantom{-}\MZ\sw s_\beta \\
0   & M_2 & \phantom{-} \MZ \cw \cos\beta  & -\MZ \cw s_\beta \\
 - \MZ \sw c_\beta &\phantom{-} \MZ \cw c_\beta  & 0 & -\mu\\
\phantom{-}\MZ\sw s_\beta& -\MZ \cw s_\beta & -\mu & 0
\end{pmatrix}\,.
\end{equation}
Here $M_1$ denotes the supersymmetry breaking bino mass. Furthermore
$\sw\equiv\sin\theta_{\mathrm{w}}$, $\cw\equiv\cos\theta_{\mathrm{w}}$
and $\theta_{\mathrm{W}}$ is the electroweak mixing angle. $M_Z$
denotes the $Z$ boson mass.


\begin{acknowledgments}
HD would like to thank Stefano Profumo for discussions on the recent
dark matter search data. PW is grateful to Anthony Hartin and Karsten
B\"u\ss{}er for helpful discussions on breamstrahlung and for
providing a GUINEA PIG beamstrahlung spectrum.  HD and JC were
supported by BMBF ``Verbundprojekt HEP-Theorie'' under the contract
0509PDE.  The work of HD was also supported by the Helmholtz Alliance
``Physics at the Terascale.''

\end{acknowledgments}

\bibliographystyle{h-physrev}

\begin{thebibliography}{99}

\bibitem{Nilles:1983ge}
  H.~P.~Nilles,
  Phys.\ Rept.\  {\bf 110} (1984) 1.

\bibitem{Martin:1997ns}
  S.~P.~Martin,
  arXiv:hep-ph/9709356.

\bibitem{Gildener:1976ih}
  E.~Gildener and S.~Weinberg,
  Phys.\ Rev.\  D {\bf 13} (1976) 3333;
E.~Gildener,
  Phys.\ Rev.\  D {\bf 14} (1976) 1667.

\bibitem{Veltman:1980mj}
  M.~J.~G.~Veltman,
  Acta Phys.\ Polon.\  B {\bf 12} (1981) 437.



\bibitem{Farrar:1978xj}
  G.~R.~Farrar and P.~Fayet,
  Phys.\ Lett.\  B {\bf 76} (1978) 575.

\bibitem{Dreiner:2005rd}
  H.~K.~Dreiner, C.~Luhn and M.~Thormeier,
  Phys.\ Rev.\  D {\bf 73} (2006) 075007
  [arXiv:hep-ph/0512163];
  H.~K.~Dreiner, C.~Luhn, H.~Murayama and M.~Thormeier,
  Nucl.\ Phys.\  B {\bf 795} (2008) 172
  [arXiv:0708.0989 [hep-ph]].

\bibitem{Lee:2010gv}
  H.~M.~Lee, S.~Raby, M.~Ratz, G.~G.~Ross, R.~Schieren, K.~Schmidt-Hoberg and P.~K.~S.~Vaudrevange,
  arXiv:1009.0905 [hep-ph].

\bibitem{Ellis:1983ew}
  J.~R.~Ellis, J.~S.~Hagelin, D.~V.~Nanopoulos, K.~A.~Olive and M.~Srednicki,
  Nucl.\ Phys.\  B {\bf 238} (1984) 453.

\bibitem{Pagels:1981ke}
  H.~Pagels and J.~R.~Primack,
  Phys.\ Rev.\ Lett.\  {\bf 48} (1982) 223.


\bibitem{Jungman:1995df}
  G.~Jungman, M.~Kamionkowski and K.~Griest,
  Phys.\ Rept.\  {\bf 267} (1996) 195
  [arXiv:hep-ph/9506380].

\bibitem{Drees:1996pk}
  M.~Drees, M.~M.~Nojiri, D.~P.~Roy and Y.~Yamada,
  Phys.\ Rev.\  D {\bf 56} (1997) 276
  [Erratum-ibid.\  D {\bf 64} (2001) 039901]
  [arXiv:hep-ph/9701219].

\bibitem{Cowsik:1972gh}
  R.~Cowsik and J.~McClelland,
  Phys.\ Rev.\ Lett.\  {\bf 29} (1972) 669.

\bibitem{Dreiner:2009ic}
  H.~K.~Dreiner, S.~Heinemeyer, O.~Kittel, U.~Langenfeld, A.~M.~Weber and G.~Weiglein,
  Eur.\ Phys.\ J.\  C {\bf 62} (2009) 547
  [arXiv:0901.3485 [hep-ph]].



\bibitem{Lee:1977ua}
  B.~W.~Lee and S.~Weinberg,
  Phys.\ Rev.\ Lett.\  {\bf 39} (1977) 165.


\bibitem{Hooper:2002nq}
  D.~Hooper and T.~Plehn,
  Phys.\ Lett.\  B {\bf 562} (2003) 18
  [arXiv:hep-ph/0212226].
 
\bibitem{Belanger:2002nr}
  G.~Belanger, F.~Boudjema, A.~Pukhov and S.~Rosier-Lees,
  arXiv:hep-ph/0212227.


\bibitem{Bottino:2002ry}
  A.~Bottino, N.~Fornengo and S.~Scopel,
  Phys.\ Rev.\  D {\bf 67} (2003) 063519
  [arXiv:hep-ph/0212379].

\bibitem{Bottino:2003iu}
  A.~Bottino, F.~Donato, N.~Fornengo and S.~Scopel,
  Phys.\ Rev.\  D {\bf 68} (2003) 043506
  [arXiv:hep-ph/0304080].


\bibitem{Bottino:2003cz}
  A.~Bottino, F.~Donato, N.~Fornengo and S.~Scopel,
  Phys.\ Rev.\  D {\bf 69} (2004) 037302
  [arXiv:hep-ph/0307303].

\bibitem{Bottino:2004qi}
  A.~Bottino, F.~Donato, N.~Fornengo and S.~Scopel,
  Phys.\ Rev.\  D {\bf 70} (2004) 015005
  [arXiv:hep-ph/0401186].

\bibitem{Bottino:2007qg}
  A.~Bottino, F.~Donato, N.~Fornengo and S.~Scopel,
  Phys.\ Rev.\  D {\bf 77} (2008) 015002
  [arXiv:0710.0553 [hep-ph]].

\bibitem{Abdallah:2003xe}
  J.~Abdallah {\it et al.}  [DELPHI Collaboration],
  Eur.\ Phys.\ J.\  C {\bf 31} (2003) 421
  [arXiv:hep-ex/0311019].

\bibitem{Heister:2003zk}
  A.~Heister {\it et al.}  [ALEPH Collaboration],
  Phys.\ Lett.\  B {\bf 583} (2004) 247.

\bibitem{Abbiendi:2003sc}
  G.~Abbiendi {\it et al.}  [OPAL Collaboration],
  Eur.\ Phys.\ J.\  C {\bf 35} (2004) 1
  [arXiv:hep-ex/0401026].


\bibitem{Amsler:2008zzb}
  C.~Amsler {\it et al.}  [Particle Data Group],
  Phys.\ Lett.\  B {\bf 667} (2008) 1.


\bibitem{Dedes:2001zia}
  A.~Dedes, H.~K.~Dreiner and P.~Richardson,
  Phys.\ Rev.\  D {\bf 65} (2001) 015001
  [arXiv:hep-ph/0106199].

\bibitem{Gogoladze:2002xp}
  I.~Gogoladze, J.~D.~Lykken, C.~Macesanu and S.~Nandi,
  Phys.\ Rev.\  D {\bf 68} (2003) 073004
  [arXiv:hep-ph/0211391].





\bibitem{Dreiner:2003wh}
  H.~K.~Dreiner, C.~Hanhart, U.~Langenfeld and D.~R.~Phillips,
  Phys.\ Rev.\  D {\bf 68} (2003) 055004
  [arXiv:hep-ph/0304289].

\bibitem{Dreiner:2007vm}
  H.~K.~Dreiner, O.~Kittel and U.~Langenfeld,
  Eur.\ Phys.\ J.\  C {\bf 54} (2008) 277
  [arXiv:hep-ph/0703009];
H.~K.~Dreiner, O.~Kittel and U.~Langenfeld,
  Phys.\ Rev.\  D {\bf 74} (2006) 115010
  [arXiv:hep-ph/0610020].

\bibitem{Kachelriess:2000dz}
  M.~Kachelriess,
  JHEP {\bf 0002} (2000) 010
  [arXiv:hep-ph/0001160].

\bibitem{Dreiner:2009er}
  H.~K.~Dreiner, S.~Grab, D.~Koschade, M.~Kramer, B.~O'Leary and U.~Langenfeld,
  Phys.\ Rev.\  D {\bf 80} (2009) 035018
  [arXiv:0905.2051 [hep-ph]].


\bibitem{Choudhury:1999tn}
  D.~Choudhury, H.~K.~Dreiner, P.~Richardson and S.~Sarkar,
  Phys.\ Rev.\  D {\bf 61} (2000) 095009
  [arXiv:hep-ph/9911365].

\bibitem{Nakamura:2010xx}
K. Nakamura et al. (Particle Data Group), J. Phys. G 37, 075021 (2010)

\bibitem{Belanger:2003wb}
  G.~Belanger, F.~Boudjema, A.~Cottrant, A.~Pukhov and S.~Rosier-Lees,
  JHEP {\bf 0403} (2004) 012
  [arXiv:hep-ph/0310037].

\bibitem{Bottino:2008xc}
  A.~Bottino, N.~Fornengo, G.~Polesello and S.~Scopel,
  Phys.\ Rev.\  D {\bf 77} (2008) 115026
  [arXiv:0801.3334 [hep-ph]].

\bibitem{Niro:2009mw}
  V.~Niro, A.~Bottino, N.~Fornengo and S.~Scopel,
  Phys.\ Rev.\  D {\bf 80} (2009) 095019
  [arXiv:0909.2348 [hep-ph]];
  S.~Scopel,
  AIP Conf.\ Proc.\  {\bf 1115} (2009) 111.

\bibitem{Badin:2009cf}
  A.~Badin, G.~K.~Yeghiyan and A.~A.~Petrov,
  arXiv:0909.5219 [Unknown].

\bibitem{Feldman:2010ke}
  D.~Feldman, Z.~Liu and P.~Nath,
  arXiv:1003.0437 [Unknown].

\bibitem{Fornengo:2010mk}
  N.~Fornengo, S.~Scopel, A.~Bottino,
  
  [arXiv:1011.4743 [hep-ph]].


\bibitem{Bernabei:2010mq}
  R.~Bernabei {\it et al.},
  Eur.\ Phys.\ J.\  C {\bf 67} (2010) 39
  [arXiv:1002.1028 [astro-ph.GA]].

\bibitem{Aalseth:2010vx}
  C.~E.~Aalseth {\it et al.}  [CoGeNT collaboration],
  arXiv:1002.4703 [astro-ph.CO].

\bibitem{Kuflik:2010ah}
  E.~Kuflik, A.~Pierce and K.~M.~Zurek,
  Phys.\ Rev.\  D {\bf 81} (2010) 111701
  [arXiv:1003.0682 [hep-ph]].

\bibitem{Das:2010ww}
  D.~Das and U.~Ellwanger,
  arXiv:1007.1151 [hep-ph].

\bibitem{Gunion:2010dy}
  J.~F.~Gunion, A.~V.~~Belikov, D.~Hooper,
  
  [arXiv:1009.2555 [hep-ph]].

\bibitem{Draper:2010ew}
  P.~Draper, T.~Liu, C.~E.~M.~Wagner {\it et al.},
  
  [arXiv:1009.3963 [hep-ph]].

\bibitem{Vasquez:2010ru}
  D.~A.~Vasquez, G.~Belanger, C.~Boehm, A.~Pukhov and J.~Silk,
  arXiv:1009.4380 [hep-ph].

\bibitem{Belikov:2010yi}
  A.~V.~Belikov, J.~F.~Gunion, D.~Hooper and T.~M.~P.~Tait,
  arXiv:1009.0549 [hep-ph].

\bibitem{Kim:2009ke}
  Y.~G.~Kim, S.~Shin,
  JHEP {\bf 0905 } (2009)  036.
  [arXiv:0901.2609 [hep-ph]].

\bibitem{Fitzpatrick:2010em}
  A.~L.~Fitzpatrick, D.~Hooper and K.~M.~Zurek,
  arXiv:1003.0014 [Unknown].

\bibitem{Andreas:2010dz}
  S.~Andreas, C.~Arina, T.~Hambye, F.~S.~Ling and M.~H.~G.~Tytgat,
  Phys.\ Rev.\  D {\bf 82} (2010) 043522
  [arXiv:1003.2595 [hep-ph]].

\bibitem{Frandsen:2010yj}
  M.~T.~Frandsen and S.~Sarkar,
  Phys.\ Rev.\ Lett.\  {\bf 105} (2010) 011301
  [arXiv:1003.4505 [hep-ph]].

\bibitem{Alves:2010dd}
  D.~S.~M.~Alves, S.~R.~Behbahani, P.~Schuster and J.~G.~Wacker,
  JHEP {\bf 1006} (2010) 113
  [arXiv:1003.4729 [hep-ph]].

\bibitem{Feldman:2010wy}
  D.~Feldman, Z.~Liu, P.~Nath and G.~Peim,
  Phys.\ Rev.\  D {\bf 81} (2010) 095017
  [arXiv:1004.0649 [hep-ph]].

\bibitem{Essig:2010ye}
  R.~Essig, J.~Kaplan, P.~Schuster and N.~Toro,
  arXiv:1004.0691 [hep-ph].

\bibitem{Chang:2010yk}
  S.~Chang, J.~Liu, A.~Pierce, N.~Weiner and I.~Yavin,
  JCAP {\bf 1008} (2010) 018
  [arXiv:1004.0697 [hep-ph]].

\bibitem{Graham:2010ca}
  P.~W.~Graham, R.~Harnik, S.~Rajendran and P.~Saraswat,
  arXiv:1004.0937 [hep-ph].

\bibitem{Kappl:2010qx}
  R.~Kappl, M.~Ratz, M.~W.~Winkler,
  
  [arXiv:1010.0553 [hep-ph]].

\bibitem{Bae:2010hr}
  K.~J.~Bae, H.~D.~Kim, S.~Shin,
  
  [arXiv:1005.5131 [hep-ph]].

\bibitem{Ahmed:2009zw}
  Z.~Ahmed {\it et al.}  [The CDMS-II Collaboration],
  Science {\bf 327} (2010) 1619
  [arXiv:0912.3592 [astro-ph.CO]].


\bibitem{Ahmed:2008eu}
  Z.~Ahmed {\it et al.}  [CDMS Collaboration],
  Phys.\ Rev.\ Lett.\  {\bf 102} (2009) 011301
  [arXiv:0802.3530 [astro-ph]].

\bibitem{Angle:2007uj}
  J.~Angle {\it et al.}  [XENON Collaboration],
  Phys.\ Rev.\ Lett.\  {\bf 100} (2008) 021303
  [arXiv:0706.0039 [astro-ph]].


\bibitem{Aprile:2010um}
  E.~Aprile {\it et al.}  [XENON100 Collaboration],
  arXiv:1005.0380 [astro-ph.CO].


\bibitem{Baudis:2010tk}
L. Baudis, talk at the 18th international conference on Supersymmetry and 
the Unification of Fundamental Interactions, Bonn, Germany, 2010.
http://susy10.uni-bonn.de/program.php


\bibitem{MoortgatPick:2008yt}
  G.~Moortgat-Pick,
  J.\ Phys.\ Conf.\ Ser.\  {\bf 110} (2008) 072027
  [arXiv:0801.2414 [hep-ph]].

\bibitem{:2007sg}
  J.~Brau {\it et al.}  [ILC Collaboration],
  arXiv:0712.1950 [physics.acc-ph].


\bibitem{Martyn:2004jc}
  H.~U.~Martyn,
  arXiv:hep-ph/0408226.

\bibitem{Martyn:2004ew}
  H.~U.~Martyn,
  arXiv:hep-ph/0406123.

\bibitem{Freitas:2004re}
  A.~Freitas, H.~U.~Martyn, U.~Nauenberg and P.~M.~Zerwas,
  arXiv:hep-ph/0409129.


\bibitem{Allanach:2002nj}
  B.~C.~Allanach {\it et al.},
in {\it Proc. of the APS/DPF/DPB Summer Study on the Future of Particle Physics (Snowmass 2001) } ed. N.~Graf,
  Eur.\ Phys.\ J.\  C {\bf 25} (2002) 113
  [arXiv:hep-ph/0202233].

\bibitem{AguilarSaavedra:2001rg}
  J.~A.~Aguilar-Saavedra {\it et al.}  [ECFA/DESY LC Physics Working Group],
  arXiv:hep-ph/0106315.


\bibitem{Djouadi:2007ik}
  G.~Aarons {\it et al.}  [ILC Collaboration],
  arXiv:0709.1893 [hep-ph].


\bibitem{Porod:2003um}
  W.~Porod,
  Comput.\ Phys.\ Commun.\  {\bf 153} (2003) 275
  [arXiv:hep-ph/0301101].


\bibitem{Bartl:1987zg}
  A.~Bartl, H.~Fraas and W.~Majerotto,
  Z.\ Phys.\  C {\bf 34}, 411 (1987).

\bibitem{Bartl:1997yi}
  A.~Bartl, H.~Eberl, S.~Kraml, W.~Majerotto, W.~Porod and A.~Sopczak,
  Z.\ Phys.\  C {\bf 76}, 549 (1997)
  [arXiv:hep-ph/9701336].

\bibitem{Blochinger:2002zw}
  C.~Blochinger, H.~Fraas, G.~A.~Moortgat-Pick and W.~Porod,
  Eur.\ Phys.\ J.\  C {\bf 24}, 297 (2002)
  [arXiv:hep-ph/0201282].

\bibitem{Kraml:1999qd}
  S.~Kraml,
  arXiv:hep-ph/9903257.


\bibitem{Schulte:1999tx}
  D.~Schulte,
  Beam-beam simulations with GUINEA-PIG, CERN-CLIC-NOTE-387.

\bibitem{ILDLOI}
  ILD Concept Group, The International Large Detector, Letter of Intent,
  March 2009,
  available at \url{http://www.ilcild.org/documents/ild-letter-of-intent}

\bibitem{calice:2008ma}
  C.~Adloff {\it et al.}  [CALICE Collaboration],
  J.\ Phys.\ Conf.\ Ser.\  {\bf 160} (2009) 012065
  [arXiv:0811.2354 [physics.ins-det]].


\bibitem{Feldman:1997qc}
  G.~J.~Feldman and R.~D.~Cousins,
  Phys.\ Rev.\  D {\bf 57}, 3873 (1998)
  [arXiv:physics/9711021].

\bibitem{Bartl:1985fk}
  A.~Bartl, H.~Fraas and W.~Majerotto,
  Z.\ Phys.\  C {\bf 30}, 441 (1986).


\bibitem{Bartl:1986hp}
  A.~Bartl, H.~Fraas and W.~Majerotto,
  Nucl.\ Phys.\  B {\bf 278} (1986) 1.



\end{thebibliography}


\end{document}